\documentclass[twocolumn]{article}

\usepackage{preprint}

\usepackage[T1]{fontenc}

\usepackage{amsmath}
\usepackage{amssymb}
\usepackage{siunitx}

\usepackage{color}
\usepackage{graphicx}
\usepackage{tikz}

\usepackage{multirow}
\usepackage{multicol}
\usepackage{float}
\usepackage{subcaption}

\usepackage{longtable}

\usepackage[inline]{enumitem}

\usepackage{natbib}

\usepackage{xwatermark}
\newwatermark[firstpage,color=gray!60,angle=90,scale=0.32, xpos=3.9in,ypos=0]{\href{https://doi.org/}{\color{gray}{Publication DOI}}}
\newwatermark[firstpage,color=gray!90,angle=0,scale=0.28, xpos=0in,ypos=-5in]{Preprint submitted to Journal of Crystal Growth}

\usepackage[colorlinks = true,
            linkcolor = blue,
            urlcolor  = blue,
            citecolor = blue,
			anchorcolor = black]{hyperref}
			
\usepackage{abstract}

\newcommand{\proj}{\mathrm{proj}}

\title{Inverse response behaviour in the bright ring radius measurement of the Czochralski process I: Investigation}

\usepackage{authblk}

\author[1]{Halima Zahra Bukhari}
\author[1]{Morten Hovd}
\author[2\thanks{Corresponding author: \texttt{jan.winkler@tu-dresden.de}}]{Jan Winkler}

\affil[1]{Department of Engineering Cybernetics, NTNU, Trondheim, Norway}
\affil[2]{Institute of Control Theory, Faculty of Electrical \& Computer Engineering, TU Dresden, Germany}

\begin{document}

\twocolumn[ 
\begin{@twocolumnfalse} 

\maketitle

\begin{abstract}	  
This is the first part of a two-article series that deals with the investigation
of the anomalous behaviour in the radius measurement signal of the Czochralski
(Cz) process and its mitigation in a feedback control system. The inverse or
anomalous behaviour is indeed a measurement signal response, which initially is opposite
to that of the expected response. This is a crucial and limiting factor in feedback
control system design. The paper presents the development of a rigorous 3D
ray-tracing method to investigate the inverse response behaviour in the
measurement signal. The results of this study provide an insight into the
dynamic behaviour of the Cz growth process. It can serve as a guideline for
achieving effective crystal radius control, which is addressed in the second part of this article series. 
\end{abstract}  

\vspace{0.35cm}

  \end{@twocolumnfalse} 
] 
\saythanks

\section{Introduction} \label{sec:Intro}
\begin{figure}[ht!]
	\centering
	\includegraphics[width=1.0\linewidth]{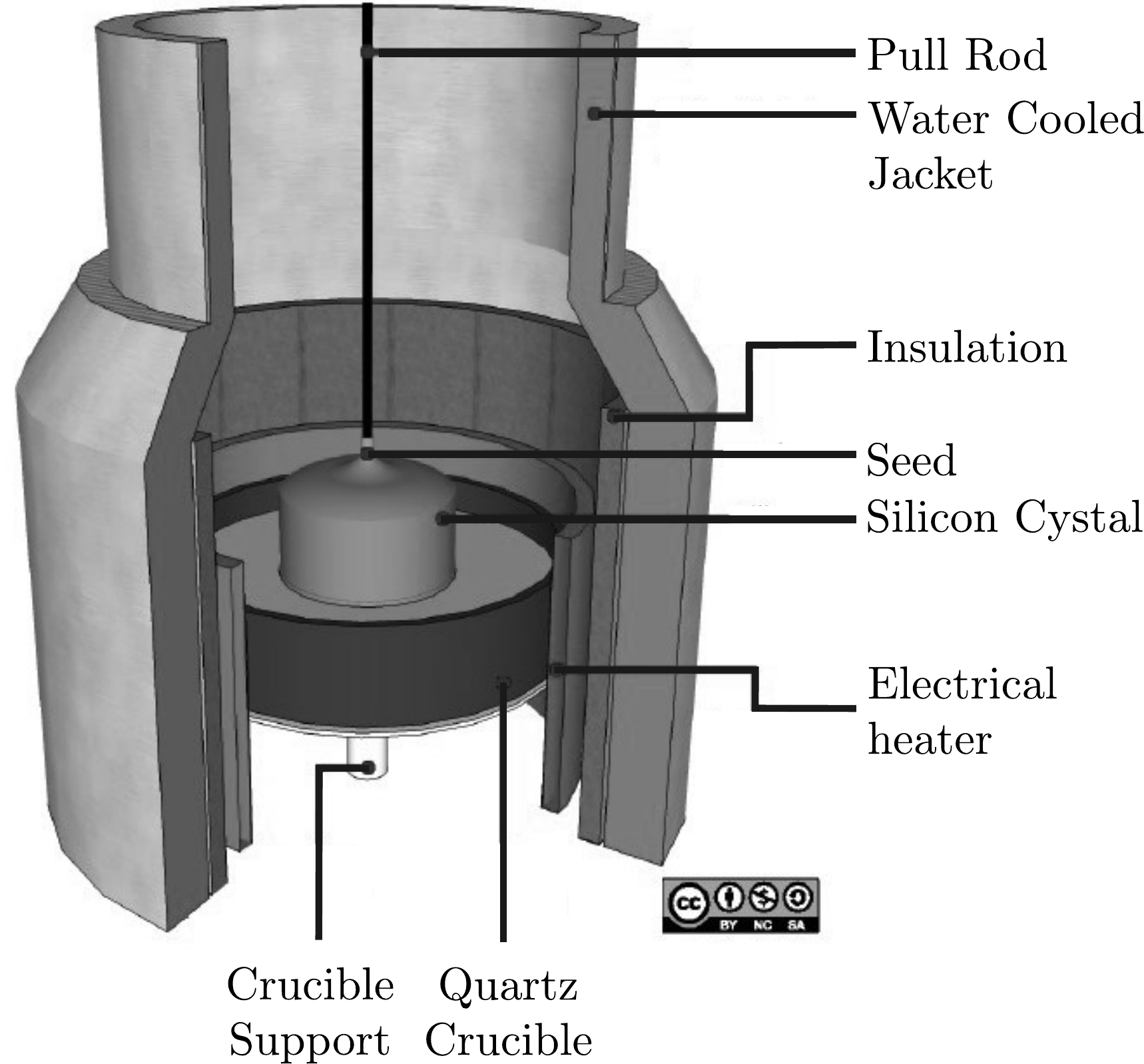}
	\caption{Assembly of a typical Cz puller \citep{rahmanpour2017model}.}
	\label{fig:furnace}
\end{figure}

\begin{figure}[ht!]
	\centering
	\includegraphics[width=\linewidth]{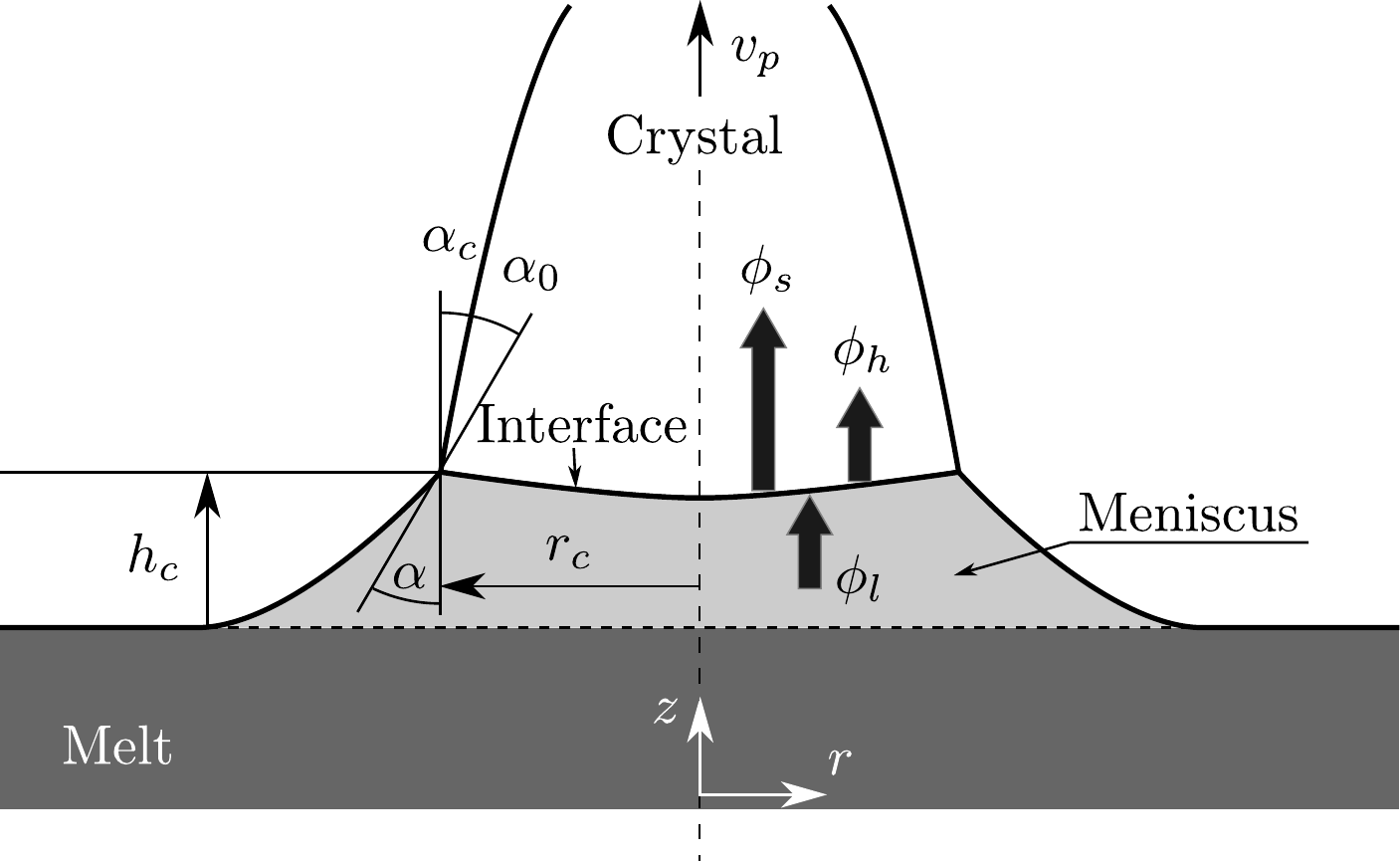}
	\caption{Schematic view of the melt-crystal interface.}	
	\label{fig:interface}
\end{figure}

The Czochralski (Cz) crystal growth process is a well-established and highly
automated method, that is indeed a workhorse for the commercial production of
monocrystalline silicon (Si) ingots. It also plays an important role in the
growth of germanium and oxide crystals. The crystallization of the
mono-crystalline ingots takes place inside a Cz puller assembly as sketched in
Fig.~\ref{fig:furnace}. In case of Si crystal growth the feed material is first
melted inside a rotatable quartz crucible. For that purpose heaters surround the
crucible from all sides, including the base. Once Si is heated up to a
temperature slightly higher than its melting point, the growth of a crystal
ingot is then initiated by immersing a seed crystal into the melt and then
gradually pulling it upwards. The pulling rod that supports the seed crystal is
not just pulled upwards but also rotated --  usually in the direction opposite to that of crucible rotation. Also the crucible itself is lifted gradually such that
the solid-liquid interface of the growing crystal is kept in a fixed position.
The pulling of a crystal from the melt results in a slightly raised liquid volume
that extends from the growing crystal interface to the flat melt surface. This volume is denoted as the \emph{meniscus} throughout this paper.
Fig.~\ref{fig:interface} shows a schematic view. The dynamics of the meniscus is
most important for the growth of the crystal.

The growth of a crystal ingot starts with a thin \emph{neck}. After achieving the desired neck size, the neck stage gradually transitions into the \emph{body} stage by going through intermediate stages of \emph{crown} and \emph{shoulder} growth. Out of different ingot segments, it is the cylindrical segment (crystal body), that is processed later by the device manufacturers.

The primary performance objective of the Cz process is to attain a uniform cross-section throughout the body length.
Variations in crystal cross-section are commonly
referred to as \textit{pinches}. Crystal structure defects, such as the number
of inclusions, nonuniform dopant distributions, etc.\,have a much higher
propensity to occur at the pinch locations \citep{tatartchenko_ch2}. Therefore,
to avoid pinch formation, a uniform crystal growth rate and cross-section is
desired. This requires precise tracking of the crystal pulling speed and the
heater temperatures such that the meniscus is influenced in a targeted
manner. Especially in the body phase the crystal diameter should remain
constant. Unfortunately, the process is highly sensitive with respect to any
disturbances acting on it. Accordingly, it is a challenging task to design a
reliable and robust automated growth control system. 

\subsection{Measurement anomaly} \label{sec:Intro:Anomaly}
\begin{figure}
	\begin{center}
		\includegraphics[width=0.6\linewidth]{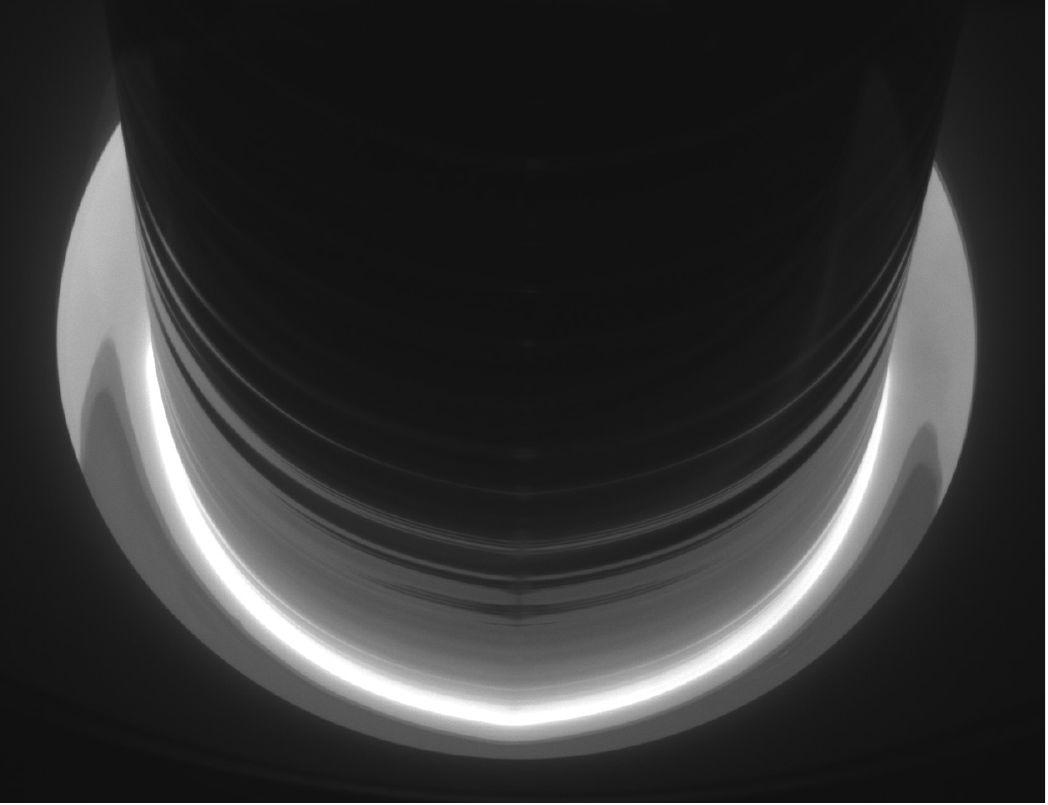}
		\caption{Actual image of the plant, captured by a CCD camera while focusing on the melt surface within the Cz growth furnace. The complete view of the bright circular ring on the meniscus is occluded by the crystal ingot in the centre. However, the radius of this bright ring serves as an estimate for the crystal radius.} 	
		\label{Fig:image_norsun}
	\end{center}
\end{figure}

Any automated feedback control system needs measured quantities in order to gain
information about the current state of the process, especially the deviation of
the quantities to be controlled from their set point values (i.e., the crystal
radius and the growth rate). In the Cz process, the measured variable generally
used for the feedback control of the crystal radius can either be the force
acting on a load cell connected with the upper end of the pulling rod (usually
referred to as \emph{weight measurement}\footnote{In fact it measures the weight of
the crystal \emph{and} the forces resulting from the surface tension and
hydrostatic pressure of the meniscus.})
\citep{levinson1959temperature,bardsley1974developments, bardsley1977weighing}
or the radius measurement of a CCD camera mounted at the top of the plant and looking downward into the vessel (Fig.~\ref{Fig:image_norsun})
\citep{patzner1967automatic,digges1975basis,lorenzini1974overview,duffar2010crystal}.
Since the boundary between crystal and the melt is indistinguishable, the
CCD-camera is adjusted to continuously focus the meniscus in the vicinity of the
three-phase boundary. It optically senses the radius of a specific bright ring
formed on the meniscus. The bright annular rings on the meniscus 
are caused by
the reflection of light by the curved meniscus in such a way that the hotter
crucible wall and the heat shield underside form varying brightness pattern on
the illuminated meniscus. Any 
clearly and consistently identifiable
point on this 
brightness
pattern, illuminated by a
specific component within the hot zone, may serve as a basis for the crystal
radius measurement.  Therefore, in the jargon of crystal growers, this very
radius measurement is termed as the \emph{bright ring radius} denoted by
$r_{br}$. Fig.~\ref{Fig:image_norsun} shows an image of the illuminated meniscus
as captured with 
a
CCD camera. It is apparent that the bright ring image is a
view-occluded glowing ring as some of the meniscus reflections from the opposite
side of the camera are obscured either by the heat shield or the cylindrical
ingot in the center or both.

\begin{figure}[ht!]
	\centering
	\includegraphics[width=1.0\linewidth]{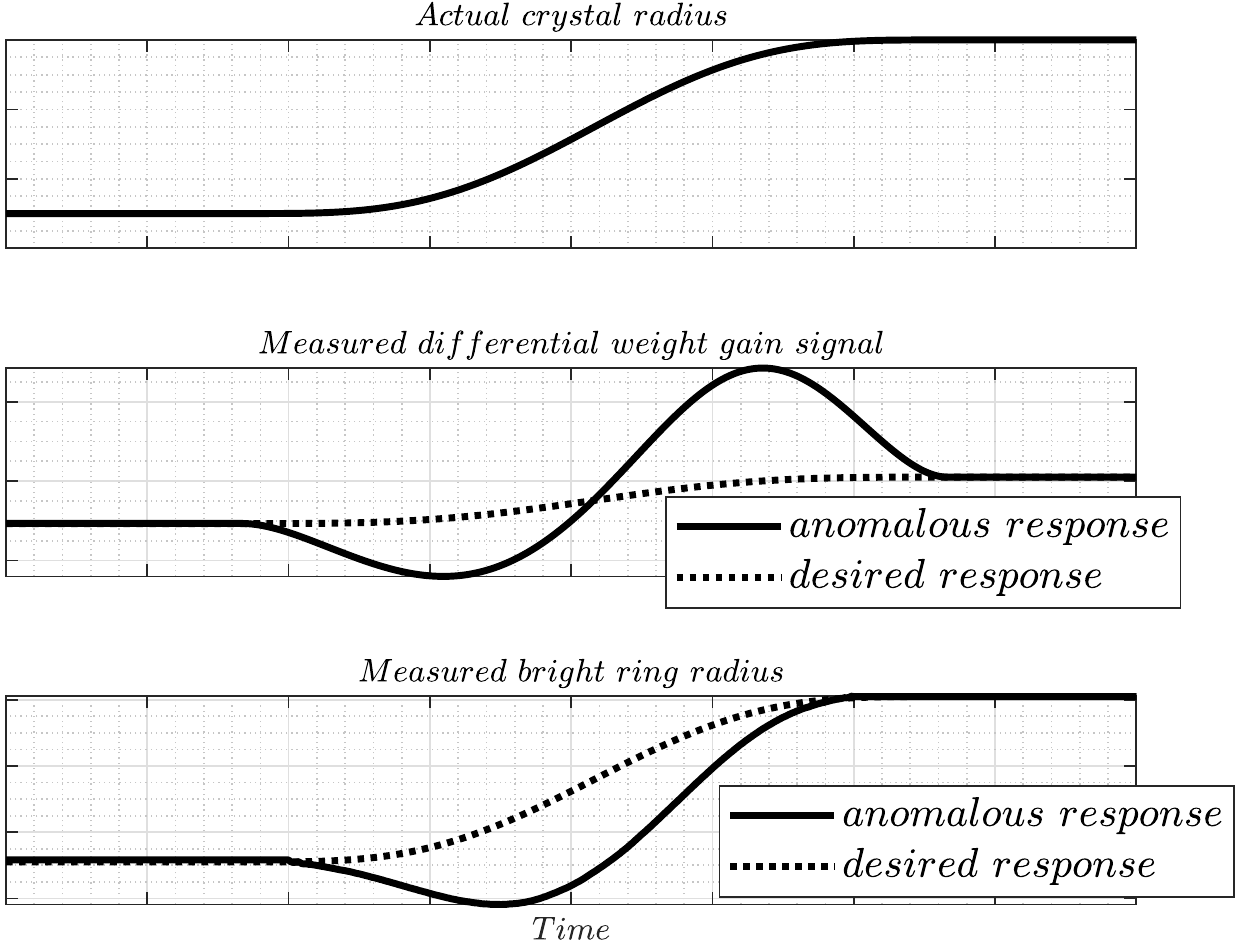}
	\caption{Illustration of anomalous behaviour in the bright ring and weight measurement signals.}
	\label{fig:anomaly_illustration}
\end{figure}

The idea behind the weight measurement method is that the change of crystal mass
per time unit divided by the pulling speed is proportional to the square of the crystal radius. For that purpose, the
measured weight signal is differentiated with respect to time and used as an
indirect value for the controlled variable. However, since the measured weight
signal is also affected by the meniscus dynamics this equality does not hold,
especially during changes of the radius. For example, in the case of a crystal
radius increase the vertical component of the meniscus' surface tension
decreases. Additionally, the melt column below the crystal (responsible for the static pressure) decreases and --
because the density of the melt is larger than that of the solid -- another
decreasing effect is added. As a consequence, the measured variable initially
responds inversely to the radius. This anomaly in
the measured signal is depicted in Fig.~\ref{fig:anomaly_illustration} (middle)
for a change of the crystal radius in a positive direction. This fact is well
known and widely investigated in the Cz crystal growth literature
\citep{bardsley1977weighingII, hurle1977control,
gevelber1988dynamics,gevelber1994dynamics}. 

A similar anomalous behaviour is known to be possessed by the bright ring
measurement in a qualitative sense \citep{gevelber1994dynamics}. For example,
with a decrease in pulling speed, the meniscus height decreases, making the
meniscus profile flatter. This effect will result in an increase in the crystal
radius, while at the same time causing the camera to initially detect a decrease
in the bright ring radius. Only after the crystal radius has grown significantly
will the camera detect an increase in the bright ring radius.
Fig.~\ref{fig:anomaly_illustration} (bottom) pictorially illustrates this
anomalous behaviour measuring the bright ring radius. Although there are
publications dealing with the calculation of the bright ring diameter
\citep{kimbel2001shape} a thorough investigation of the impact of the anomalous
behaviour on control system performance remains. This work, in particular, aims
at investigating the aforementioned from both systematic and quantitative
standpoint.

For this purpose, the first part of this two-article series contains a detailed investigation of
the phenomena causing the anomalous behaviour. The second part of the series
describes the control design to mitigate the detrimental effects of the
anomalous measurement behaviour on the control of crystal radius. Some of the
results of this two-article series have previously been presented in an abridged
form for to a control audience in conference publications
\citep{BUKHARI2019129,BUKHARI2020}. This two-article series aims to present the
results in full detail and accessible also to a crystal growth audience, without
requiring expert knowledge of control.

\subsection{Paper organization} \label{sec:Intro:Organization}
Section\,\ref{sec:Instrumentation_control} gives an overview of the control of
the Czoch\-ralski process as it is currently done in an industrial environment.
A simplified process model describing the overall Cz dynamics appropriate for
control design is derived in Section\,\ref{sec:complete_model} comprising a
rigorous description of the crystal growth dynamics (section\,\ref{sec:Cz_growth_model}) and simplified temperature dynamics (section\,\ref{sec:temperature dynamics}). A method based on ray-tracing for the
bright ring radius estimation and inverse response investigation is presented in
section\,\ref{sec:measurement_model}. Finally, Section\,\ref{sec:conclusions}
provides conclusions and points to further work to be presented in part II of this article series.

\section{Conventional control of the Cz process using optical diameter measurement} \label{sec:Instrumentation_control}
The challenging task in control system design 
is that the Cz process is a complex process with a combination of both faster as
well as slower dynamics. The dynamics at the crystallization interface
controlled by the pulling speed is quite fast. It can affect the crystal
radius, meniscus height and the corresponding growth conditions quickly compared to the slower dynamics associated with the heater
temperature input and the highly complex nonlinear heat transfer phenomena. The
heat transfer from the heaters to the crystallization front undergoes both lag and long
time delays. Thus, the effect of heat input on the crystal growth rate is
noticeable after a significant time lapse.

\begin{figure}[ht!]
\centering
	\includegraphics[width=1\linewidth]{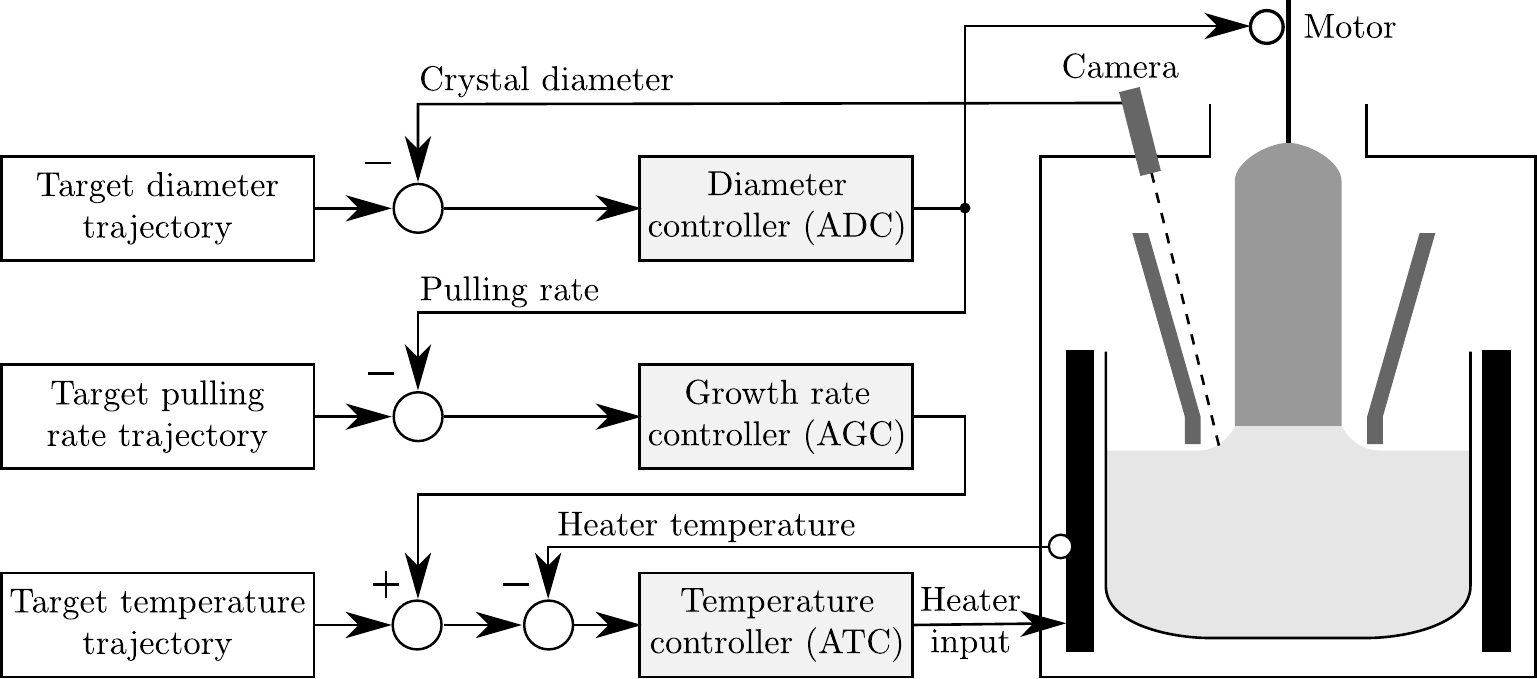}
	\caption{Schematics showing a conventional control structure of the Cz system.} 	
	\label{Fig:conventional_control_structure}
\end{figure}

In conventional Cz control these different dynamics are addressed by a cascaded controller structure comprising three control loops: The automatic
diameter controller (ADC), the  automatic growth rate controller (AGC) and the
automatic temperature controller (ATC), respectively \citep{lee2005mpc},
(cf.\,Fig.~\ref{Fig:conventional_control_structure}).

During a typical growth cycle in the body stage, a target temperature trajectory
is applied to the temperature controller.
In an actual process, the target temperature trajectory has an
increasing trend
to compensate for the following events
occurring throughout the growth cycle within the Cz growth chamber: 
\begin{enumerate*}[label=(\roman*)]
	\item A gradual uplift of the crucible, therefore progressively reducing the
	crucible exposure to the heaters.
	\item With the ongoing crystallization, the crystal continues to protrude
	into the colder areas above the heat shield, thereby increasing the heat
	transfer away from the interface.
\end{enumerate*}
The temperature controller takes its measurement from a pyrometer. There is a small opening in the containment structure that allows for pyrometer insertion, providing the pyrometer with a view of 
graphite lining surrounding the heater element. This thereby
provides a measurement for the heater temperature.

The growth rate controller and the temperature controller are connected in
series with each other. The tracking error for the pulling speed triggers the growth rate
controller, which in turn, adds a trim value to the target temperature
trajectory. Due to sluggish dynamics from heater power to crystal growth rate, a time-varying target temperature trajectory is introduced. The intent of introducing the target temperature trajectory is to counteract the factors described above that change the heat transfer characteristics, and thereby, provide anticipative action to reduce variations in the crystal growth rate.
A well-designed temperature trajectory implies better tracking of pulling speed and will therefore reduce the contribution from the growth rate controller. 


\section{A low order model of the Cz dynamics}\label{sec:complete_model}
For a rigorous investigation of the bright ring anomaly and its impact on the control system performance, a basic model of the Cz dynamics is required. The Czochralski process comprises of crystal growth dynamics at the interface explained in detail in Section\,\ref{sec:Cz_growth_model} and additionally of
the heater/ temperature dynamics which significantly affects the outcome of
the process. Especially the growth rate $v_g$ driving the basic growth dynamics
is a result of the thermal situation at the crystallization interface. However,
an accurate model representing the temperature dynamics and the complex heat
transfers within the Cz process would have to be derived from describing the
system using partial differential equations, and thereafter applying
discretization on a fine grid to arrive at a large set of ordinary differential
equations representing the temperature dynamics \cite{Dornberger1996,Hoffmann2003}. While such a model may be
appropriate for process design studies, it is commonly considered too large and
impractical for control design studies. In Section\,\ref{sec:temperature dynamics},
we will therefore develop two simplified models instead, based on a coarse lumped
model, adapted from the model presented in \citep{rahmanpour2017model}. While
the heat transport is of little importance for the investigation of the bright
ring measurement in Section\,\ref{sec:anomalous_behaviour} of this paper, it is
central when evaluating control performance in a \emph{qualitatively} reasonable manner
in part II of this article series. Therefore, Section\,\ref{sec:temperature
dynamics} can be skipped by readers interested only in the analysis of the
bright ring measurement.

\subsection{Growth dynamics}\label{sec:Cz_growth_model}
The standard Cz growth model commonly referred to as either the
hydromechanical-geometrical model \citep{winkler2010czochralski} or simply the
crystal growth dynamics at the crystal-melt solidification interface, is given
by (cf. Fig.~\ref{fig:interface} also):
\begin{subequations}\label{basic_dynamics}
	\begin{align}
	\dot{r}_c & = v_g\tan (\alpha_c) \\
	\dot{h}_c & = v_p - v_g\label{h_dot}\\
	r_{br}&=f_{br}(r_c,h_c)\\
	\alpha_c& = \arcsin \bigg\{1-\bigg(\frac{h_c}{a}\bigg)^2\bigg[1+0.6915\bigg(\frac{r_c}{a}\bigg)^{-1.1}\bigg]\bigg\}-\alpha_0\label{alpha_c_johansen}\\
	v_g &= \frac{\phi_s-\phi_l}{\rho_s \Delta H}\label{eq:vg_basic}
	\end{align}
\end{subequations}
where $r_c$ is the crystal radius, $h_c$ is the height of the meniscus at the
three-phase boundary, $v_p$ is the pulling speed, $v_g$ is the growth rate of
the crystal (in axial direction assuming a flat solid-liquid interface) and $\alpha_c$ is the cone angle at the
interface. Finally, the output $r_{br}$ is the bright ring radius obtained from
the camera image. Its dependence on the crystal radius and the meniscus height can be expressed in general by the function $f_{br}$. That this approach is reasonable will be shown later in this paper. With reference to (\ref{eq:vg_basic}), $\phi_s$ is the
heat flux from the interface into the crystal, while $\phi_l$ is the heat flux
from the meniscus to the interface. $\rho_s$ is the density of the solid
crystal, and $\Delta H$ is the specific latent heat of fusion. The derivative of
the meniscus height expressed in (\ref{h_dot}) assumes perfect compensation for
melt level changes through crucible lift.

The expression of $\alpha_c$ in (\ref{alpha_c_johansen}) is derived from the
analytical approximation of meniscus height $h_c$ given by
\cite{johansen1994improved}. The overall growth angle is  expressed as $\alpha
= \alpha_0 + \alpha_c$, where $\alpha_0$ is the contact angle at constant
radius growth, i.e., $\alpha_c = 0$. We will assume $\alpha_0 = \SI{11}{\degree}$
\citep{tatarchenko1993shaped,RAHMANPOUR2017353}. The Laplace constant $a$, also termed
the \emph{capillary length}, is
given by $a=\sqrt{\frac{2\sigma_{LG}}{\rho_{L}\cdot g}}$ with the specific
surface tension $\sigma_{LG}$ of the Si melt, the Si melt mass density $\rho_L$
and gravitational acceleration $g$. One has a value of
$a=\SI{7.62}{\milli\meter}$ as calculated for Si with
$\sigma_{LG}=\SI{0.732}{\newton\per\metre}$ and
$\rho_{L}=\SI{2570}{\kilogram\per\meter\cubed}$

Under the steady-state growth conditions, the heat flux $\phi_l$ entering the
interface from the meniscus, the heat flux $\phi_h = \varrho_s \Delta H
v_g$ released due to phase change and the heat flux $\phi_s$ directed into the
crystal are balanced. The crystal growth rate \eqref{eq:vg_basic} is defined on
the basis of this heat balance. The heat balance across the crystallization
interface should be maintained such that the net flow of heat is towards the
crystallizing interface. In other words, the continuous heat loss from the meniscus into the crystal (heat of fusion/latent heat) ensures the ongoing crystallization/solidification \citep{winkler2010czochralski}.

\subsection{Heater/Temperature dynamics}\label{sec:temperature dynamics} 
\begin{figure}[ht!]
	\begin{subfigure}{\linewidth}
		\centering
		\includegraphics[width=0.9\linewidth]{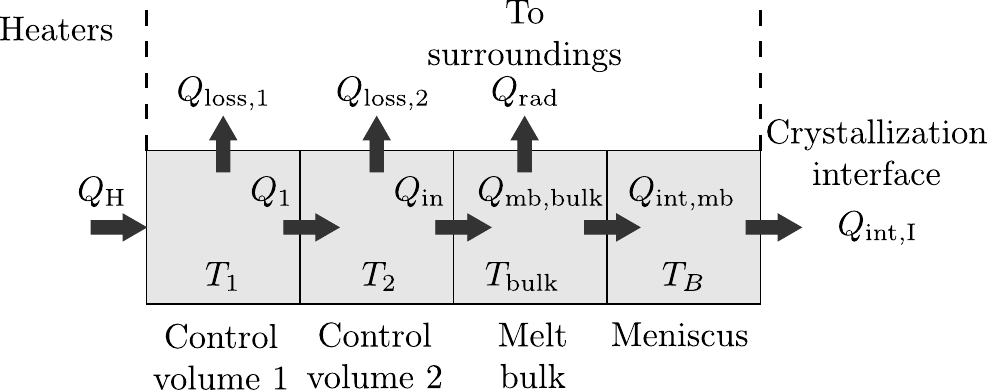}\\
		\caption{Model I: Conduction based heat transport across the meniscus.}
		\label{fig:heat_transport_mod1}
	\end{subfigure}\\
	\begin{subfigure}{\linewidth}
		\centering
		\includegraphics[width=0.75\linewidth]{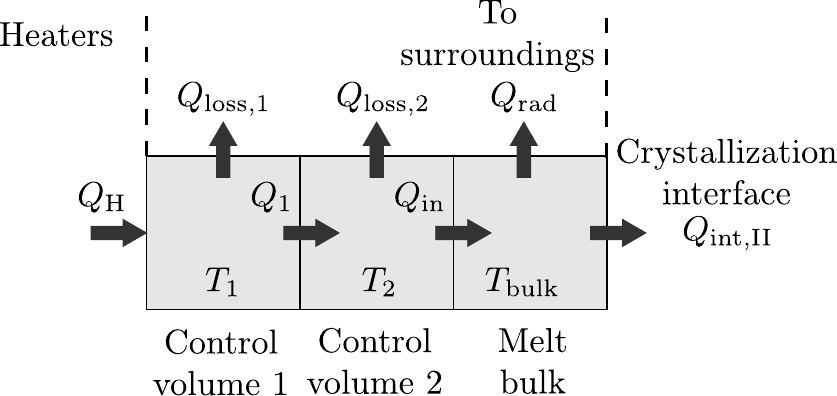}
		\caption{Model II: Convection based heat transport across the meniscus.}
		\label{fig:heat_transport_mod2}
	\end{subfigure}
	\caption{Illustration of the lumped heat transport models with their control volumes.}
	\label{fig:heat_transport_mod}
\end{figure}

The total path for the heat transfer, from the heaters to the interface, is
divided into control volumes. A control volume is a fictitious volume with
constant physical properties. For each control volume a heat balance can be
established from which an ordinary differential equation for its temperature can
be deduced. In the following, two different assumptions are used as a basis for
describing the mode of heat transfer and resulting in an expression for $\phi_l$
in (\ref{eq:vg_basic}).  
\begin{enumerate}[label=\Roman*., widest=II]
\item The heat transfer from the bulk of the melt into the meniscus is caused by
convection, while the heat transfer across the meniscus itself is based on pure
conduction. This results in a model consisting of four control volumes (cf.\,Fig.\,\ref{fig:heat_transport_mod1}). In this model, the heat flux entering the solid-liquid interface from the melt will depend on the meniscus height, as will be shown in Section\,\ref{sec:heatmodel1}.
\item The heat transfer from the bulk of the melt to the interface is based on
convection only resulting in a model consisting of three control volumes (cf.\,Fig.\,\ref{fig:heat_transport_mod2}). Here, the varying meniscus height does not have any influence on the heat flux entering the solid-liquid interface (cf.\,section\,\ref{sec:heatmodel2}).
\end{enumerate}

Though these two assumptions are in a sense two extremes, however, the actual heat transfer across the meniscus is likely to be a combination of both convection and conduction. Thus, a control design that works for both models can be expected to work for the heat transfer mode(s) occurring in the actual process.

As in our previous work \citep{BUKHARI2019129,BUKHARI2020}, we will refer to the resulting models as Model I and Model II, respectively. 

\subsubsection{Heat transport in Model I}\label{sec:heatmodel1}
In the case of Model I (cf.\,Fig.\,\ref{fig:heat_transport_mod1}), the first two control volumes are used to provide a coarse second order approximation to the dynamics of the heat transfer from the heater to the melt.
The temperature $T_1$ of control volume 1 may roughly correspond to the temperature measured by the pyrometer on the graphite lining surrounding the heaters, while the second control volume of fictional temperature $T_2$ introduces an additional time lag representing the thermal inertia of the inner assemblies surrounding the crucible. Their dynamics is given by
\begin{subequations} \label{eq:T1ThDot}
\begin{align}
\dot{T}_1 &= \frac{Q_{H}-Q_{1}-Q_{loss,1}}{\tau_1}\label{eq:T_1_dot} \\
\dot{T}_2 &= \frac{Q_{1}-Q_{in}-Q_{loss,2}}{\tau_2} \label{eq:T_h_dot}
\end{align}
\end{subequations}
where $\tau_1, \tau_2$, represent the parameters proportional to time-delay for the heat transfers through control volumes 1 and 2, respectively \cite{RAHMANPOUR2017353}. These parameters are chosen to approximate an effective time delay of \SI{600}{\second} for each of the two control volumes.
 In \eqref{eq:T1ThDot} one has the heat input $Q_{H}$ from the
heaters, i.e., the manipulated variable for the ATC controller, the thermal
energy $Q_{1}$ that enters the control volume 2 from control volume 1 and the
thermal energy $Q_{in}$ that enters the melt. These heat transfer rates are modelled using \cite{RAHMANPOUR2017353}

\begin{equation}\label{eq:Q1}
Q_{1} = \beta_1\,(T_1-T_2)
\end{equation}
and
\begin{equation}\label{eq:Qin}
Q_{in} = \beta_2\,(T_2-T_{bulk}),
\end{equation}
with the overall heat transfer coefficients $\beta_1, \beta_2$ between the
control volumes  and the melt bulk with temperature $T_{bulk}$. The two heat losses
$Q_{loss,1}$ and $Q_{loss,2}$ in \eqref{eq:T1ThDot} are assumed to be constant
for a shorter time scale of dynamical analysis. 

The last two control volumes to the right of this model (cf.\,Fig.\,\ref{fig:heat_transport_mod1}) distinctively represent the
melt bulk region and the meniscus, respectively. The heat transfer is modelled as two heat transfers in series: the \emph{convective}
heat transfer from the bulk of the melt into the meniscus and the \emph{conductive} heat transfer across the meniscus
to the crystallization interface. The dynamics of
the melt bulk temperature $T_{bulk}$ reads
\begin{equation}\label{eq:T_s_dot}
\dot{T}_{bulk} = \frac{Q_{in} - Q_{int,I} -  Q_{rad}}{V_s\cdot\rho_l\cdot C_p},
\end{equation}
where $V_s$, $\rho_l$ and $C_p$ define the melt volume, the density and the specific heat capacity of liquid Si, respectively. The convection based heat flow
$Q_{mb,\,bulk}$ from the melt bulk region into the meniscus of temperature
$T_B$ can be calculated from
\begin{align}\label{qmb_bulk}
Q_{mb,\,bulk} = \beta_{conv,I}\,(T_{bulk}-T_{B}),
\end{align}
with the overall heat transfer coefficient $\beta_{conv,I}$.

The conductive heat flow $Q_{int,\,mb}$ from the meniscus to the crystallization interface (with Si melting point temperature $T_S$) reads
\begin{equation}\label{qint_mb}
	Q_{int,\,mb} = \frac{k_{cond,I}\,A_{i}\,(T_{B}-T_{S})}{h_c},
\end{equation}
where $A_{i}\, =\,\pi\,r_c^2$ is the cross-sectional area of the solidification interface and $k_{cond,I}$ is the heat conductivity of liquid Si\footnote{As can be seen in Eq.\,\eqref{qint_mb}, the temperature gradient is assumed to be $(T_{B}-T_{S})/h_c$ which is a quite rough but common approximation in lumped parameter models of the Cz process \cite{hurle1990dynamics}. In reality, the thickness of the thermal boundary layer is the driving force for conductive heat transfer. But since this layer is not modelled here this approximation is used. It simply reflects the heuristic assumption that the closer the interface to the hot melt, the more the crystallization is inhibited \cite{neubert2014nonlinear}.}. Due to the
short height of the meniscus it is reasonable -- on the timescale of relevance
for crystal growth -- to neglect the \emph{dynamics} of $T_B$, i.e., $\dot{T}_B = 0$, and assume the two heat transfers $Q_{int,\,mb}$ and $Q_{mb,\,bulk}$ to be equal. 
This allows us to eliminate $T_B$ from \eqref{qmb_bulk},\eqref{qint_mb} and arrive at the following expression for the overall heat flow $Q_{int,I}$
entering the crystallization interface from the bulk:
\begin{equation}\label{Qint1}
Q_{int,I}=\beta_{int}(T_{bulk}-T_S)
\end{equation}
The coefficient $\beta_{int}=(\beta_{conv,I}^{-1}\,+\,h_c\,k_{cond,I}^{-1}\,A_{i}^{-1})^{-1}$ is
the overall heat transfer coefficient that combines the two coefficients from expressions \eqref{qmb_bulk} and \eqref{qint_mb} in series. Similarly (from \eqref{qmb_bulk},\eqref{qint_mb}) the heat transfer coefficient $\beta_{conv,I} $ is given by
\begin{equation}\label{hconv_I}
\beta_{conv,I} = \frac{k_{cond,I}\,(T_{B,0}-T_S)\,A_{i}}{h_c\,(T_{bulk,0}-T_{B,0})},
\end{equation}
using initial steady-state values $T_{B,0}, T_{bulk,0}$ for $T_{B}, T_{bulk}$, respectively.
Hence, its value is assumed constant throughout the simulations. Moreover, $\beta_{conv,I}$ is adjusted to achieve the observed crystal growth rate.

Finally, the radiative heat loss $Q_{rad}$ from the melt surface is expressed as:
\begin{equation}\label{Qrad}
Q_{rad}=A_{fm}\,F_{mc}\,\epsilon_m\,\sigma\,(T_{bulk}^4-T_{env}^4)
\end{equation}
where $A_{fm}$ is the free melt surface area expressed as $A_{fm} =
\pi\,(R_{cru}^2-r_c^2)$, $F_{mc}$ is the radiation view factor considering the
heat radiation from the free melt surface to the crystal surroundings,
$\epsilon_m$ is the melt emissivity, $\sigma$ is the \emph{Stefan-Boltzmann
constant} and $T_{env}$ is the temperature of the environment.

\subsubsection{Heat transport in Model II}\label{sec:heatmodel2}
In case of model II convective heat transport from the melt to the
crystallization interface is considered only, i.e., the bulk of the melt, as well as
the temperature of the meniscus, are at the same temperature level. Hence, the
model consists of only three control volumes
(cf.~Fig.~\ref{fig:heat_transport_mod2}). 

The temperature dynamics of the first two control volumes is given by \eqref{eq:T1ThDot}, while the bulk melt temperature $T_{bulk}$ is given by:
\begin{equation}\label{eq:T_s_dotII}
	\dot{T}_{bulk} = \frac{Q_{in} - Q_{int,II} - Q_{rad}}{V_s\cdot\rho_l\cdot C_p}.
\end{equation}
In the above equation \eqref{eq:T_s_dotII}, the radiative heat loss $Q_{rad}$ is formulated as in Eq.\,\eqref{Qrad}, while $Q_{int,II}$, the \emph{convective} heat transfer from the melt bulk to the crystallization interface, is given as
\begin{equation}
	Q_{int,II} = \beta_{conv,II}\,(T_{bulk}-T_{S}),
\end{equation}
where $\beta_{conv,II}$ represents the convective heat transfer from the bulk of the melt to the crystallization interface. The value of $\beta_{conv,II}$ given by 
\begin{equation*}
\beta_{conv,II}=\frac{\phi_l A_i}{(T_{bulk,0}-T_S)}
\end{equation*}
is adjusted to achieve the observed crystal growth rate.

\subsection{Overall model}\label{sec:OverallModel}
With the results from sections\,\ref{sec:Cz_growth_model} and \ref{sec:temperature dynamics} the overall Cz dynamics, including both growth and temperature dynamics, can be written in the so-called state-space form with the state $\mathbf{x} = (r_c, h_c, T_{bulk}, T_1, T_2)^T$:
\begin{subequations}\label{eq:temp_model}
	\begin{align}
	\dot{\textbf{x}}\,&=\begin{pmatrix}
	v_g\tan (\alpha_c) \\
	v_p - v_g\\
	(Q_{in}-Q_{int}-Q_{rad})/(V_s\,\rho_l\,C_p)\\
	(Q_{H}-Q_{1}-Q_{loss,1})/\tau\\
	(Q_{1}-Q_{in}-Q_{loss,2})/\tau
	\end{pmatrix}=\textbf{f(x,u)} \label{eq:Overall:f}\\
	\mathbf{y}&=
	\begin{pmatrix}
	r_{br}\\
	T_1
	\end{pmatrix},\qquad \mathbf{u} = 
	\begin{pmatrix}
	v_{p}\\
	Q_{H}
	\end{pmatrix}.
	\end{align}
\end{subequations}
In \eqref{eq:Overall:f}, $Q_{int}$ can either be $Q_{int,I}$ or $Q_{int,II}$ depending on the
choice of the heater model. The growth rate $v_g$ is calculated according to
\eqref{eq:vg_basic} with $\phi_l = Q_{int}/A_i$. In this model, $u$ indicates the input
vector comprising of two manipulating inputs ($v_p,\,Q_H$), while the
measured output $y$ comprises of the bright ring radius $r_{br}$ and the temperature $T_1$
sensed by the pyrometer. Note, that in case of model I the growth rate depends on the meniscus height (cf. (\ref{qint_mb}), (\ref{Qint1})). A method for determining $r_{br}$, proposed in the
next section, is the main topic of this paper.

\section{Ray-tracing method for bright ring radius estimation}\label{sec:measurement_model}
In this section a rigorous ray-tracing simulation, combined with the crystal
growth dynamics, is developed to simulate the camera image of the illuminated
meniscus. The simulated camera image is then used to calculate the bright ring
radius for control system design and analysis. In \cite{kimbel2001shape}, 
ray-tracing is used to estimate the bias between the actual crystal radius and the measured bright ring radius for a static case. This paper presents a method to simulate the actual camera image and the dynamic analysis of the resultant
bright ring radius measurement that can aid in effective and improved control system design. 

The light incident on the meniscus from different components in the hot zone assembly gets reflected from the meniscus surface and captured by the camera as a bright ring image. Fig.~\ref{fig:raytracingsetup2} shows a simplified
ray-tracing setup. It presents the vertical cross-sectional view of the Cz growth furnace (crucible wall, heat-shield, and camera location). For the sake of simplicity, only one ray from each source is shown to be incident onto the meniscus and reflected thereof before reaching the camera. Though the rays may undergo multiple reflections before reaching the camera, an instance of the ray reflected twice from the meniscus is shown by a dashed line in the same figure. 

It is clear that a key factor when modeling the dynamics of the bright ring radius is the knowledge of the meniscus shape. The meniscus shape can be calculated from the so-called \emph{Laplace-Young equation} that accounts for surface tension, gravity, and hydrostatic pressure to express the shape and height of the meniscus. Unfortunately, there is no analytical solution to the Laplace-Young equation, which, therefore, has to be solved numerically \citep{huh1969shapes}. An alternative to the aforementioned approach is the use of an
analytical approximation of the meniscus shape.

\subsection{Meniscus shape approximation}\label{sec:meniscus_shape}
An approximation of the meniscus profile based on the actual crystal radius $r_c$ at the interface and the meniscus height $h_c$ is presented in \cite{hurle1983analytical}:
\begin{equation}\label{hurle_men_h1}
\begin{split}
r(h_c,r_c,z) &= r_c + \sqrt{\frac{2}{A}-h_c^2}-\sqrt{\frac{2}{A}-z^2}\\
&-\frac{1}{\sqrt{2A}}\ln\bigg[\frac{z}{h_c}\cdot\frac{\sqrt{2}+\sqrt{2-A \cdot h_c^2}}{\sqrt{2}+\sqrt{2-A\cdot z^2}}\bigg]
\end{split}
\end{equation}
where $r$ and $z$ are the radial and vertical coordinates of the meniscus
surface, respectively. Thus, one has $(r,z) = (r_c,h_c)$ where the meniscus
connects to the crystal, while roughly $(r,z) = (R_{cru},0)$ at the crucible wall. The
parameter $A$ is defined as
\begin{equation*}
A=\frac{1}{a^2}+\frac{\cos(\alpha)}{2r_c\,h_c}.
\end{equation*} 
This meniscus profile can be extended to define a full 3D meniscus surface by
rotating the profile about the $z$ axis, i.e., $360^{\circ}$ along the azimuthal plane.

The ray-tracing simulation relies primarily on the knowledge of the Cz growth
model parameters $(r_c, h_c)$ as discussed in Section\,\ref{sec:Cz_growth_model} combined with the information about the meniscus
shape profile \eqref{hurle_men_h1} to simulate the image of the CCD camera.

\subsection{Hot zone geometry}
\begin{figure}[ht!]
	\centering
	\includegraphics[width=\linewidth]{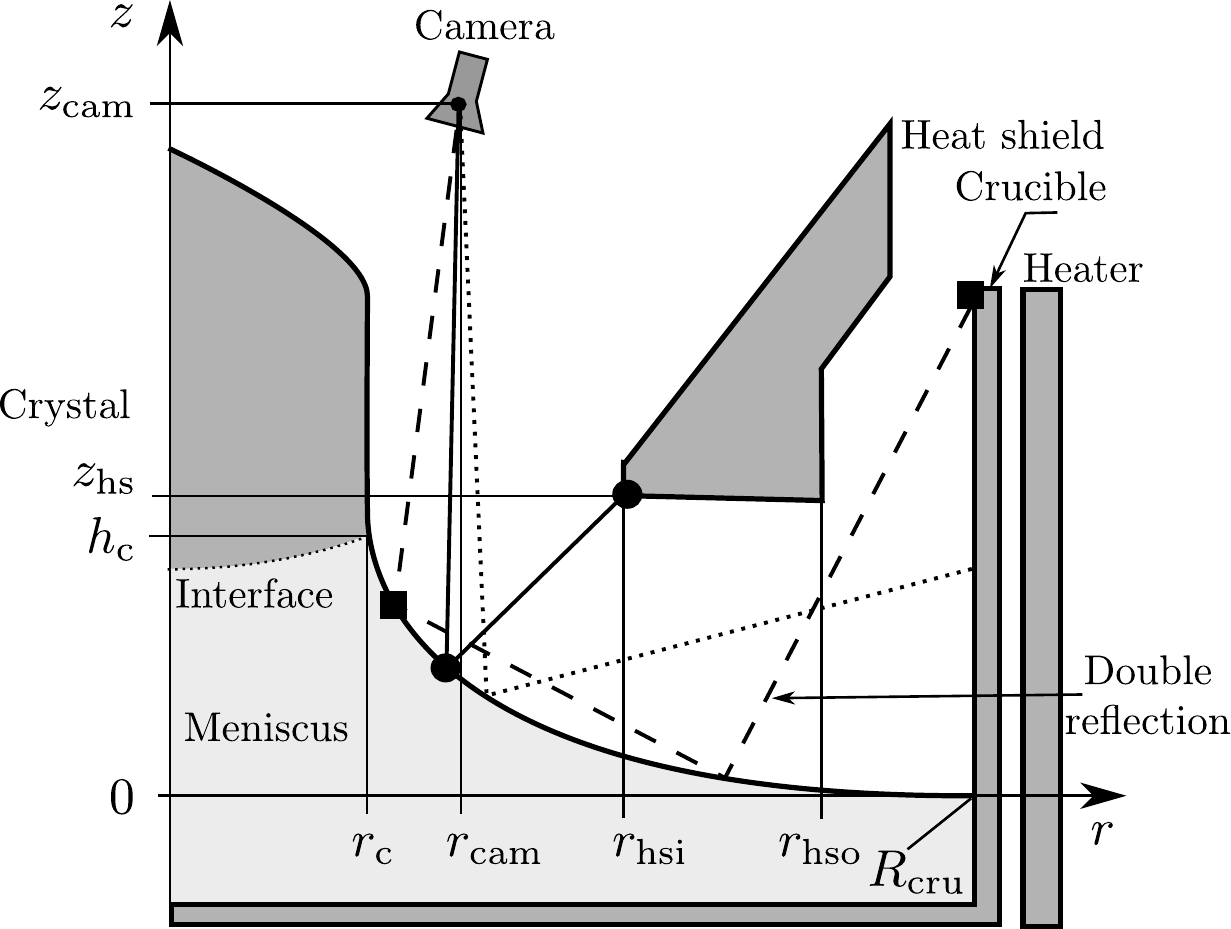}
	\caption{Ray-tracing set up showing incident and reflected light rays within the growth furnace.}
	\label{fig:raytracingsetup2}
\end{figure}

\begin{figure}[ht!]
	\centering
	\includegraphics[width=0.9\linewidth]{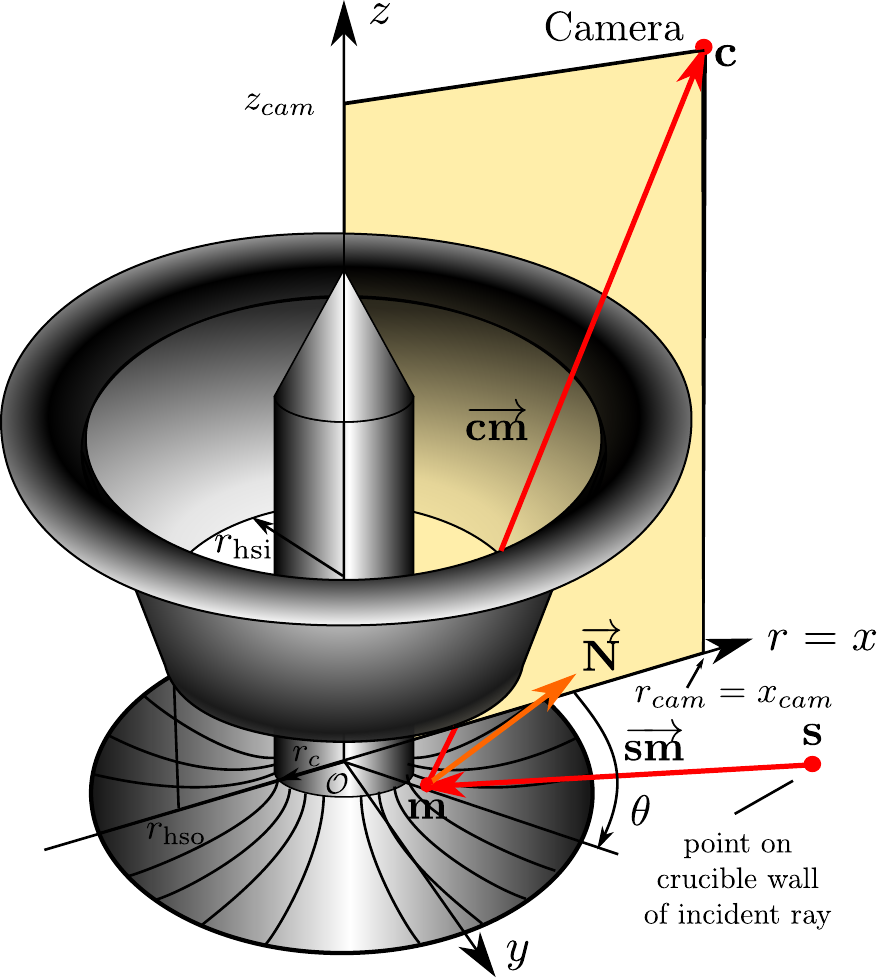}
	\caption{3D ray-tracing scheme featuring an instance of an incident ray $\vec{\textbf{sm}}$ emerging from the crucible wall. The reflected ray from a point ($\textbf{m}$) reaches the camera with location marked as ($\textbf{c}$). The incident and reflected rays may or may not exist in the same plane.}
	\label{fig:cz_3D_scheme}
\end{figure}

The aspects of the hot zone geometry used in describing the
ray-tracing method are shown in Figs.~\ref{fig:raytracingsetup2} and \ref{fig:cz_3D_scheme}. 
The reference frame origin $\mathcal{O}$ is placed at the level of the free melt surface such that the horizontal $x$-axis and the lateral $y$-axis form the $xy$-plane spanning the
base of the meniscus, while the $z$-axis extends vertically along the center of
the ingot. The camera is located in the $xz$-plane ($y=0$). In terms of
cylindrical coordinates, the camera location is in the $rz$-plane, i.e., in the
$0^\circ$ azimuthal plane. The hot zone components, such as the crucible wall and
annular heat shield surrounding the growing crystal, have dimensions defined as:
\begin{itemize}[labelindent=1ex,leftmargin=*]
	\item  Crucible radius:$\,\,R_{cru}$
	\item The coordinates for the heat shield underside are described in terms
	of its height as well as inner and outer radii given by:
	\begin{itemize}[leftmargin=*]
	\item $\,\,z_{hs}$ is the height of the heat shield underside w.r.t. the free melt surface
	\item $\,\,r_{hsi}$ is the inner radius of the heat shield underside w.r.t. $\mathcal{O}$
	\item $\,\,r_{hso}$ is the outer radius of the heat shield underside w.r.t. $\mathcal{O}$
	\end{itemize}
	\item The camera location w.r.t. $\mathcal{O}$ is described by the position
	vector $\vec{\textbf{p}}_c$ such that:
	$\vec{\textbf{p}}_c=x_{cam}\hat{i}\,+\, 0\hat{j}\,+\, z_{cam}\hat{k}$
	\footnote{$\hat{i}$, $\hat{j}$ and $\hat{k}$ are the unit vectors directed
	along $x$, $y$ and $z$-axes, respectively}, where
	\begin{itemize}[leftmargin=*]
		\item $\,\,z_{cam}$ is the height of the camera w.r.t. the free melt surface
		\item $x_{cam}$ is the radial location of the camera, i.e.,
		$x_{cam}=r_{cam}$ as $y_{cam}=0$
	\end{itemize}		
\end{itemize}

\subsection{Computation of tangents and normals to the meniscus surface}
Followed by the generation of 3D meniscus surface is the calculation of tangents
and normals to the entire meniscus surface. This, in turn, helps to determine
the incoming and outgoing rays for the camera image simulation. The tangents and
the unit normals to the meniscus surface are mathematically notated by
$\vec{\textbf{T}}$ and $\vec{\textbf{N}}$ in $\mathbb{R}^3$, respectively. Since
the meniscus surface is axisymmetric, tangents and normals can be calculated using the 2D meniscus curve, and then rotate these around the $z$-axis to the required azimuthal orientation. 

For any arbitrary point
$(x_{m_0}, 0, z_{m_0})$ (or equivalently
($r_{m_0}, 0, z_{m_0}$) in cylindrical coordinates) on the meniscus profile such that $z_{m_0} = f(r_{m_0})$, 
the tangent vector can be calculated as: 
\begin{equation}
\begin{split}
\vec{\textbf{T}}|_{(r_{m_0},0,z_{m_0})} &=\frac{1}{\sqrt{1+f^{\prime 2}(r)}} \begin{bmatrix}
\frac{\Delta r}{\Delta r}\\
\frac{\Delta \theta}{\Delta r}\\
\frac{\Delta f(r)}{\Delta r}
\end{bmatrix}\Bigg|_{(r_{m_0},0,f(r_{m_0})}\\
&= \frac{1}{\sqrt{1+f^{\prime 2}(r_{m_0})}}\begin{bmatrix}
1\\
0\\
f^{\prime}(r_{m_0})
\end{bmatrix}.
\end{split}
\end{equation} 
The normal vector is orthogonal to the tangent, and hence we have
\begin{equation}
	\vec{\textbf{N}}|_{(r_{m_0},0,z_{m_0})}=\frac{1}{\sqrt{1+f^{\prime 2}(r_{m0})}}\begin{bmatrix}
	-f^{\prime}(r_{m_0})\\
	0\\
	1
	\end{bmatrix}.
\end{equation} 
Finally, the unit normal to any arbitrary point in question (say $\textbf{m}_{(r,\theta,z)}$) can be determined by rotating $\vec{\textbf{N}}|_{(r_{m_0},0,z_{m_0})}$
through an angle $\theta$ about the $z$-axis, such that 
\begin{equation}\label{eq:unit_normal}
\vec{\textbf{N}}_{m_{(r_{m},\theta,z_{m})}} = R_z(\theta)\cdot\vec{\textbf{N}}|_{(r_{m_0},0,z_{m_0})},
\end{equation} 
where $R_z(\theta)$, the rotation matrix yielding the desired rotation through an arbitrary angle $\theta$, about the $z$-axis is given as
\begin{equation}
R_z(\theta)=\begin{bmatrix}
\mathrm{cos}(\theta)&-\mathrm{sin}(\theta)&0\\
\mathrm{sin}(\theta)&-\mathrm{cos}(\theta)&0\\
0&0&1
\end{bmatrix}.
\end{equation}

\subsection{Computing incoming and outgoing rays}
A rigorous ray-tracing method that abides the laws of reflection is employed to
simulate the bright ring formation over the curved meniscus.
Fig.~\ref{fig:cz_3D_scheme} illustrates the schematic of the 3D ray-tracing. The
outgoing ray that enters the camera after reflection from a point \textbf{m}
on the meniscus (described by position vector $\vec{\textbf{p}}_m$ in cylindrical coordinates as $(r_m,\theta,z_m)$  or
as $(x_m,y_m,z_m)$ in Cartesian coordinates) is given by
\begin{multline}
\overrightarrow{\textbf{cm}}=\vec{\textbf{p}}_c-\vec{\textbf{p}}_m=\\(x_{cam}-x_m)\hat{i}\,-\,y_m\hat{j}\,+\,(z_{cam}-z_m)\hat{k}.
\end{multline} 
Based on the knowledge of meniscus normals given in (\ref{eq:unit_normal}), the
incoming and outgoing rays to the meniscus and camera, respectively can be
determined such that the angle of incidence equals the angle of reflection,
though the two angles need not to be be co-planar. For a specific point on
the meniscus, the projection of the reflected ray $\overrightarrow{\textbf{cm}}$
in the direction of normal vector $\vec{\textbf{N}}$ at the same point is given
by
\begin{equation}
    \proj_{\vec{\textbf{N}}}\overrightarrow{\textbf{cm}} = ( \overrightarrow{\textbf{cm}}\cdot\vec{\textbf{N}})\vec{\textbf{N}},
\end{equation}
where $(\cdot)$ indicates the vector dot product operation. Similarly, the
projection of $\overrightarrow{\textbf{cm}}$ in the direction orthogonal to
$\vec{\textbf{N}}$ is given as 
\begin{equation}
 \proj_{\perp\longrightarrow \vec{\textbf{N}}}\overrightarrow{\textbf{cm}}=\overrightarrow{\textbf{cm}}-\proj_{\vec{\textbf{N}}}\overrightarrow{\textbf{cm}}=\overrightarrow{\textbf{cm}}-(\overrightarrow{\textbf{cm}}\cdot\vec{\textbf{N}})\vec{\textbf{N}}.
\end{equation}
The orthogonal projection of incoming ray vector $\overrightarrow{\textbf{sm}}$
is the same as the reflected  ray vector $\overrightarrow{\textbf{cm}}$ but with
the opposite sign, i.e.,
$\proj_{\vec{\textbf{N}}}\overrightarrow{\textbf{sm}}=-\overrightarrow{\textbf{cm}}+(\overrightarrow{\textbf{cm}}\cdot\vec{\textbf{N}})\vec{\textbf{N}}$
However, the projection of incoming ray vector in the direction of normal vector
$\vec{\textbf{N}}$ is the same as that of reflected ray vector, i.e.,
$\proj_{\vec{\textbf{N}}}\overrightarrow{\textbf{sm}}=(\overrightarrow{\textbf{cm}}\cdot\vec{\textbf{N}})\vec{\textbf{N}}$
Thus, the incoming ray from a given source point inside the hot zone,
represented as $\overrightarrow{\textbf{sm}}$, is given by
\begin{equation}
\begin{split}
\overrightarrow{\textbf{sm}} &= ( \overrightarrow{\textbf{cm}}\cdot\vec{\textbf{N}})\vec{\textbf{N}} - \overrightarrow{\textbf{cm}}+(\overrightarrow{\textbf{cm}}\cdot\vec{\textbf{N}})\vec{\textbf{N}}\\ 
&=2(\overrightarrow{\textbf{cm}}\cdot\vec{\textbf{N}})\vec{\textbf{N}}-\overrightarrow{\textbf{cm}}\\
& = 2\vec{\textbf{N}}(\vec{\textbf{N}}^T\, \overrightarrow{\textbf{cm}})-\overrightarrow{\textbf{cm}}\\
&=(2\vec{\textbf{N}}\vec{\textbf{N}}^T-\textbf{I}) \overrightarrow{\textbf{cm}},
\end{split}
\end{equation}
where $I$ is the identity matrix. It is noteworthy that the incident rays
$\overrightarrow{\textbf{sm}}$ and the reflected rays
$\overrightarrow{\textbf{cm}}$ may pass through planes corresponding to different azimuthal orientations. In order to determine if the source of illumination on the illuminated meniscus is the crucible wall or the heat shield
underside, the following steps can be followed: 

The sum of the two vectors, given by
	\begin{equation}\label{eq:ps_pm_sm}
	\vec{\textbf{p}}_s = \vec{\textbf{p}}_m\,+\,\overrightarrow{\textbf{sm}},
	\end{equation}
	describes a position vector $\vec{\textbf{p}}_s$ for the point of incidence,
	\textbf{s} w.r.t. $\mathcal{O}$. In \eqref{eq:ps_pm_sm}, $\overrightarrow{\textbf{sm}}$ shall not be confused with the incoming ray (cf. Fig.~\ref{fig:cz_3D_scheme}). Instead, the expression $\eqref{eq:ps_pm_sm}$ makes use of vector mathematics (\emph{head-to-tail rule of vector addition}) by interpreting $\overrightarrow{\textbf{sm}}$ as a vector that has to be extended appropriately up to the source point, as in \eqref{eq:cruc_wall}.

	The intersection of the incident ray with the crucible wall can be
		found by scaling the position vector $\,\,\overrightarrow{\textbf{sm}}$ by a factor
		`$k$' such that it emerges from the crucible wall of radius $R_{cru}$.
		This can be achieved by solving the expression given in
		(\ref{eq:cruc_wall}) for the positive root of `$k$':
	\begin{equation}\label{eq:cruc_wall}
	(x_m+k\,\overrightarrow{\textbf{sm}}_x)^2+(y_m+k\,\overrightarrow{\textbf{sm}}_y)^2=R_{cru}^2.
	\end{equation}

	The elevation of the scaled up incident ray
	$k\,\overrightarrow{\textbf{sm}}$ emerging from the crucible wall is given
	by its $k\, \overrightarrow{\textbf{sm}}_z$.  The double reflection (shown in Fig.~\ref{fig:raytracingsetup2})
		is caused by the incoming ray that emerges from the portion of the
		crucible wall lying below the melt level, i.e., the $z$-coordinate of
		$k\,\overrightarrow{\textbf{sm}}$ is negative ($k\,
		\overrightarrow{\textbf{sm}}_z<0$).
		 The origin of rays undergoing double reflection can be found by calculating another reflection where the incoming ray hits the meniscus surface.  The details are omitted for brevity.  Note that it is possible for some rays to be reflected more than twice, especially when the meniscus close to the crystal is highly curved, which may occur when the meniscus is high.  Such multiple reflections have not been considered further in this work -- but the calculations required to include them in the ray tracing are more tedious than difficult. 

	The incident ray from the heat shield is the one whose $x$ and
	$y$-coordinates at $z_{hs}$ satisfy the following inequality condition:
	\begin{equation}
	r_{hsi}<\sqrt{sm^2_x + sm^2_y}<r_{hso}:
	\end{equation}

Thus, a 3D ray-tracing simulation, based on the procedure outlined above, traces
every ray that reaches the camera back to its emission point (the complete
annular heat shield surrounding the growing crystal and the cylindrical crucible
wall containing the molten Si). The rays, which are reflected twice from the
meniscus, are also included in the simulated bright ring image. However, some of
the reflections from the meniscus surface, lying on the side farther to the
camera, are obscured either due to the presence of a cylindrical crystal ingot
or by the heat shield.  Likewise, the heat shield blocks many of the light rays
which emanate from various emission points, above and beyond the heat shield
underside, from reaching the camera. 

\begin{figure}[ht!]
	\centering
	\includegraphics[width=1.0\linewidth]{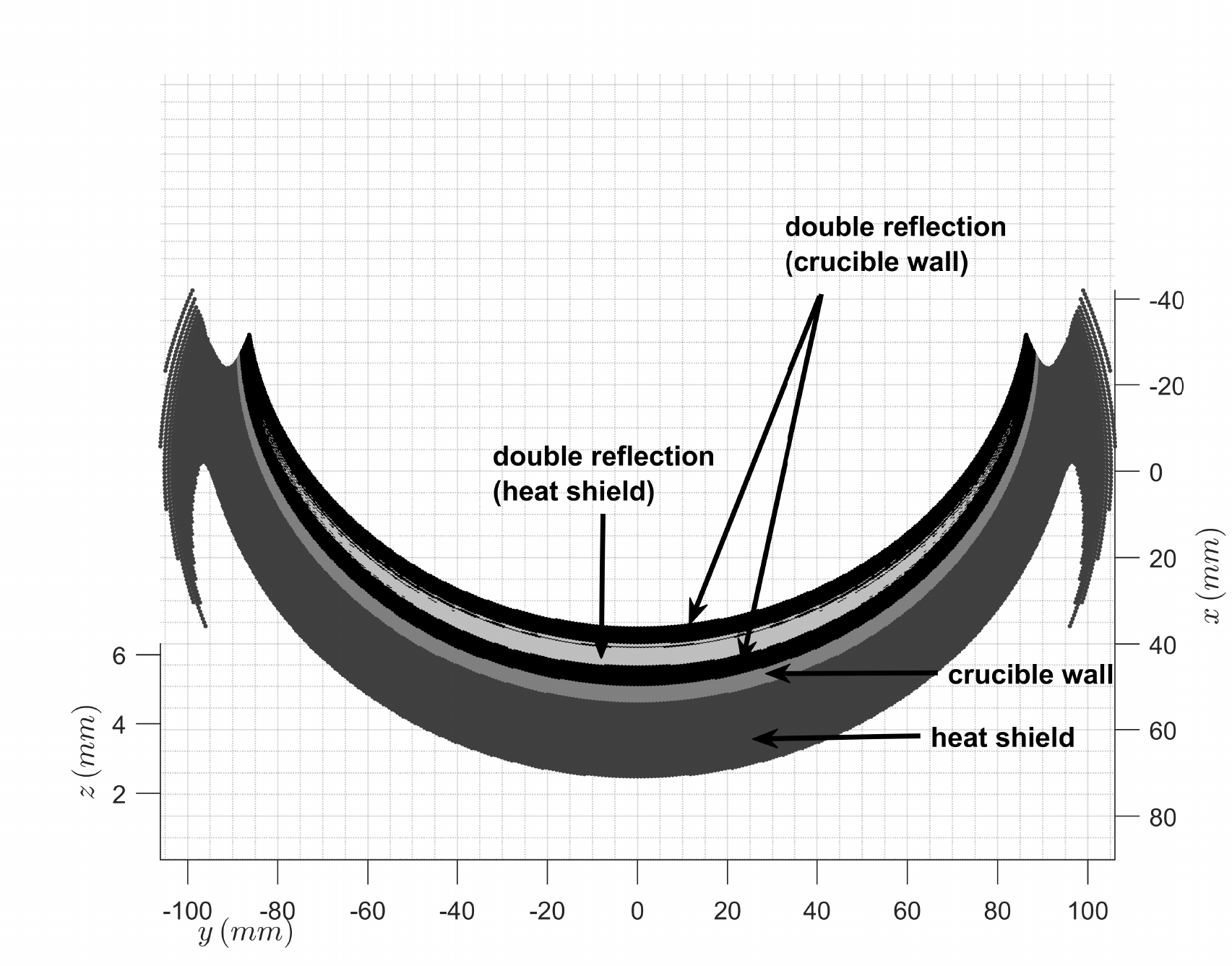}
	\caption{Meniscus image showing different regions illuminated by various components in the hot zone. The view is symmetric on either side of the $xz$-plane/camera plane\protect\footnotemark.}
	\label{fig:3D_menprofile_hotzone}
\end{figure}

The various sources of illumination on the bright ring image are depicted in
Fig.~\ref{fig:3D_menprofile_hotzone} by different gray-scale values.

\begin{figure}[ht!]
	\begin{center}
		\includegraphics[width=1.0\linewidth]{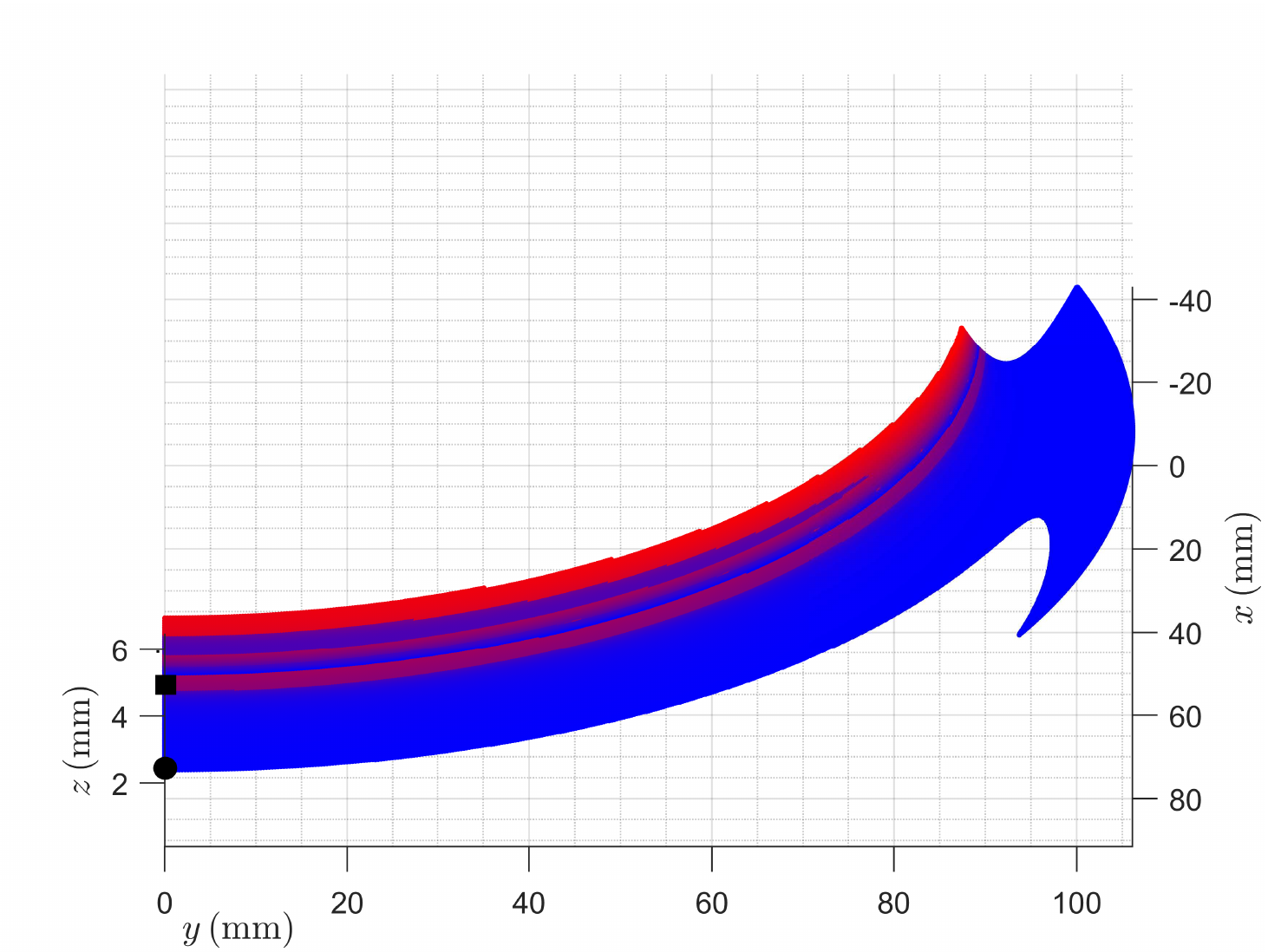}
		\caption{Theoretically calculated brightness profile as observed by the camera. Minimum brightness (blue); maximum brightness (red). Since, the view is symmetric on either side of the $xz$-plane/camera plane, the left portion of the meniscus is not shown.}
		\label{Fig:3D_brightness_profile}
	\end{center}
\end{figure}

\footnotetext{The $x$ and $y$ axes in Figs.~(\ref{fig:3D_menprofile_hotzone} $\&$ \ref{Fig:3D_brightness_profile}) define the radial coordinates of the meniscus, i.e., $r = \sqrt{x^2 + y^2}$, while $z$-axis represents the height of the meniscus above the melt surface. Thus, the plane of the camera expressed in cylindrical coordinates is $(r,\theta=0,z)$}

The brightness sensed by the camera will depend on three factors:
\begin{enumerate*}[label=\roman*.)]
\item \label{it:brightsensed:surf}The brightness of the emitting surface.
\item \label{it:brightsensed:relsurf}The orientation of the emitting surface relative to the direction of the emitted ray.
\item \label{it:brightsensed:curve}The focusing of light caused by the curvature of the meniscus.
\end{enumerate*}

Accounting accurately for \ref{it:brightsensed:surf} will require knowledge both
of the emissivity of the heat shield and crucible wall, as well as the
temperature distribution along these surfaces. Such information is not
available to the present authors, and would require the output from some very
detailed simulators. Instead, it is assumed that both the underside of the heat
shield and the crucible wall have uniform (and the same) brightness. Luckily,
this simplification does not impede our ability to study the bright ring
anomaly, as will become apparent.

Factors \ref{it:brightsensed:relsurf} and \ref{it:brightsensed:curve} are
accounted for by performing small perturbations around the point on the meniscus
where the ray is reflected before entering the camera. Let these perturbations
define the vertices of a region on the meniscus surface, and let $A_m$ be the
area of that region when projected in the direction of the ray $\overrightarrow{\textbf{cm}}$. Reflection
calculations are then performed to find the point of origin for each of the
perturbed rays. The origins of the perturbed and reflected rays define a region
on the emitting surface. Let $A_s$ be the area of this region of the emitting
surface, when projected in the direction of the emitting ray $\overrightarrow{\textbf{sm}}$. A relative
brightness measure\footnote{Note that it would be easy to account also for the
brightness of the emitting surface, if such information is available.} is then
found from the ratio of $A_s$ to $A_m$.   

The calculated brightness profile obtained, therefore, is illustrated in Fig.~\ref{Fig:3D_brightness_profile} where the color denotes the brightness of the reflection. 

In order to use a feature of the camera image for control of the crystal
radius, two obvious criteria must be fulfilled:
\begin{enumerate*}[label=\roman*.)]
\item the feature should be located close to the actual crystal radius, and
\item the feature should be clearly and reliably identifiable in the
camera image for all conditions that are expected during the body stage of the process
(i.e., for all values of crystal radius and meniscus height that are likely to
occur in the body stage).
\end{enumerate*}
Studying the calculated reflection images (and comparing to the camera image in
Fig.~\ref{Fig:image_norsun}), two such features can be identified:

\begin{itemize}[labelindent=1ex,leftmargin=*]
	\item  The highest point on the crucible wall (same as the outermost edge of
	the underside of heat shield) illuminating a point `\tikz{\draw[black,very
	thick,fill=black] (0,0) rectangle (0.125,0.125)}' on the meniscus in
	Figs.~\ref{fig:raytracingsetup2} $\&$ \ref{fig:3D_menprofile_hotzone}. 
	\item  The innermost edge of the underside of heat shield) illuminating a
	point`\tikz{\draw[black,very thick,fill=black] (0,0) circle (.5ex)}' on the
	meniscus in Figs. \ref{fig:raytracingsetup2} $\&$ \ref{fig:3D_menprofile_hotzone}.
\end{itemize} 

The first of these features indicated by `\tikz{\draw[black,very
thick,fill=black] (0,0) rectangle (0.125,0.125)}' is closer to the actual
crystal radius and is therefore the preferred feature to use for crystal radius
control. It corresponds to the lower brightness border in the overexposed Fig.\,\ref{Fig:image_norsun}. Knowledge of the point from where the light that causes this feature
originates, allows us to study the behaviour of the corresponding measurement
under dynamical process conditions.

\subsection{Anomaly detection via 3D ray-tracing simulation} \label{sec:anomalous_behaviour}
Under normal operating conditions, the physical systems rarely encounter any
abrupt changes in their physical parameters/state variables. Therefore, the
objective is to input a smooth crystal radius change to the ray-tracing
simulation and investigate how the resultant bright ring measurement differs
from the actual input signal (crystal radius). Thus, a pulling speed profile is
selected\footnote{Details can be seen in \citep{winkler2010nonlinear}} such that
it drives the Cz dynamics to generate an output that comprises of smooth profile for the crystal radius. Furthermore, it is worth mentioning that for this particular choice of a smooth pulling speed profile, the crystal growth rate is assumed to be constant. 

\begin{figure}[ht!]
	\centering
	\includegraphics[width=1\linewidth]{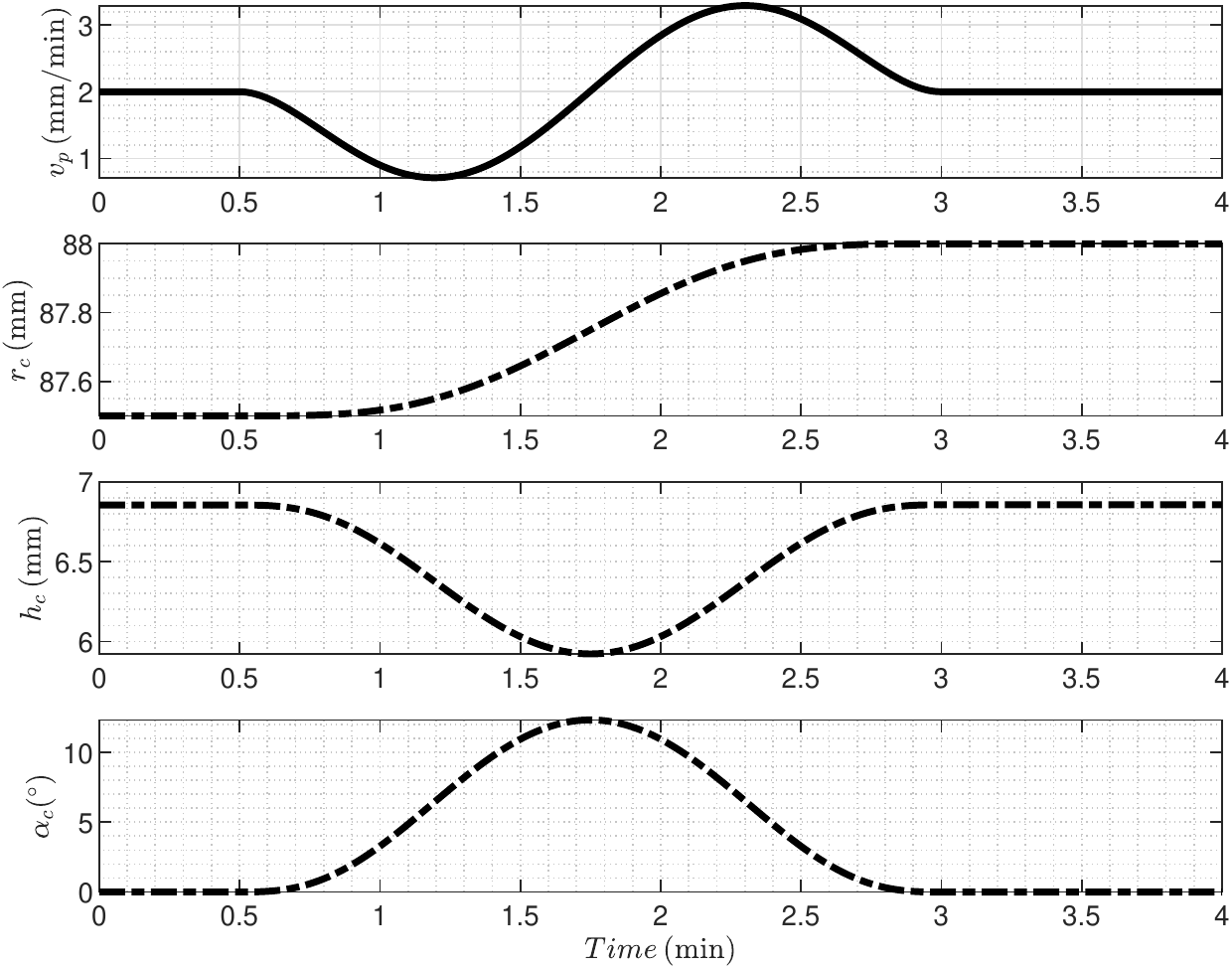}
	\caption{Smooth and continuous profile for the applied pulling speed (solid) and the resultant profiles (dash-dotted) for crystal radius, meniscus height and growth angle.}
	\label{fig:vp_profile1}
\end{figure}

Fig.~\ref{fig:vp_profile1} depicts how the chosen pulling speed profile, driving
the Cz dynamics, results in the desired crystal radius ($r_c$) variation.
Besides, the same figure shows the system trajectories for meniscus height
($h_c$) and cone angle ($\alpha_c$). 

For a smoothly varying crystal radius profile  (cf.~second subfigure in
Fig.~\ref{fig:vp_profile1}), the corresponding bright ring measurement based on
the ray-tracing method is carried out at various points along the highest
contrast line on the 3D meniscus image. One of the aforementioned illuminated
meniscus points lies in the plane of the camera ($0^\circ $ azimuth), while the
others lie in the planes at azimuthal orientations $10^\circ$, $20^\circ$,
$30^\circ$ and $40^\circ$ off the camera plane. 

\begin{figure}[H]
	\centering
	\includegraphics[width=1.0\linewidth]{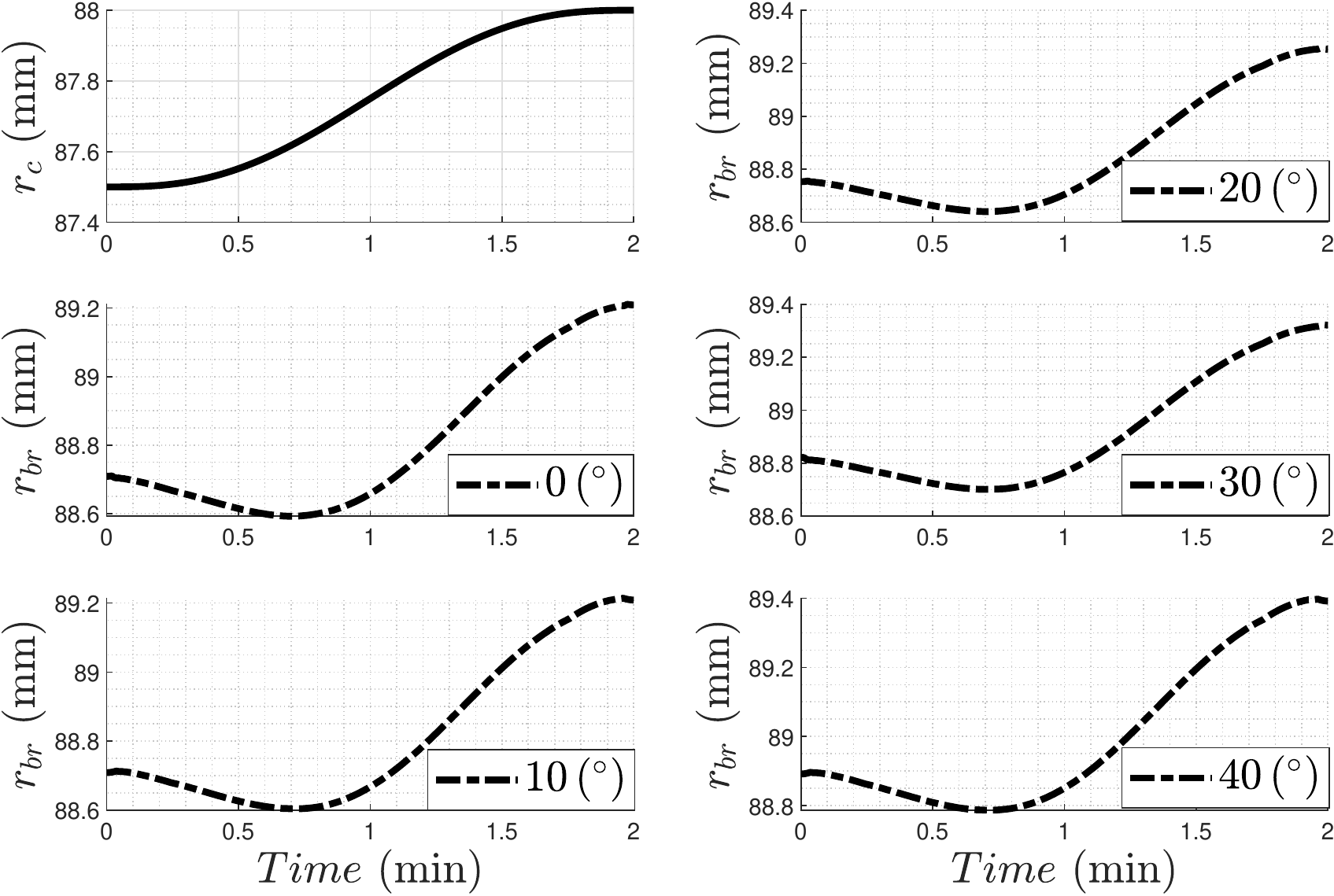}
	\caption{Actual crystal radius $r_c$ (solid) v.s. bright ring signal $r_{br}$ (dotted) measured at different azimuthal orientations, indicated respectively, at the bottom right corner of each subfigure.}
	\label{fig:3D_anomaly}
\end{figure}

The resultant bright ring radii responses versus the expected crystal radius $r_c$
response presented in Fig.~\ref{fig:3D_anomaly}, clearly reveal the presence of
the inverse response behaviour in the measurement signal. 

\section{Conclusions and way forward} \label{sec:conclusions}
This work primarily focuses on a 3D ray-tracing method that simulates the
glowing meniscus image captured with the CCD camera. The simulated camera image
provides a reference point for the bright ring radius measurement, a crucial
measure for the controlled variable. Through dynamic simulation based on the
ray-tracing scheme, the exhibition of inverse response behaviour by the bright
ring measurement signal is verified. This peculiar behaviour can pose
fundamental limitations to the design of the Cz control system. The mitigation
of this inverse response in the context of control, circumventing fundamental
limitations with feedback control by combining feedback and parallel
compensation,will extensively be dealt with in the second part of the
two-article series.

\section*{Acknowledgements}
This work has been funded by the Norwegian Research Council's ASICO project No. 256806/O20.

\bibliographystyle{plain}
\bibliography{czochralski_final}	

\onecolumn

\begin{longtable}{ |p{0.07\textwidth}|p{0.1\textwidth}|p{0.5\textwidth}|p{0.1\textwidth}| }
	\hline
	\textbf{Symbol}& \textbf{Unit} & \textbf{Description} &\textbf{Initial value}\\		
	\hline
	\endhead
	\multicolumn{4}{|c|}{\textbf{States/ parameters common between the two thermal models}} \\
	\hline
	$Q_H$&\SI{}{\kilo\watt}  &Heater input&58.6\\
	\hline
	$Q_1$&\SI{}{\kilo\watt}& Heat energy entering the control volume 2 (quartz crucible) from control volume 1 (graphite crucible)  &27.3\\
	\hline
	$Q_{in}$&\SI{}{\kilo\watt}& Heat entering the melt   &7.3\\
	\hline
	$Q_{loss,1}$&\SI{}{\kilo\watt}& Heat loss from control volume 1  &30\\
	\hline
	$Q_{loss,2}$&\SI{}{\kilo\watt}& Heat loss from control volume 2  &20\\
	\hline
	$Q_{out}$&\SI{}{\kilo\watt}& Heat transferred away from the melt  &7.3\\
	\hline
	$T_1$&\SI{}{\kelvin}& Intermediate temperature &1970\\
	\hline
	$T_2$&\SI{}{\kelvin}& Intermediate temperature &1914\\
	\hline
	$T_{S}$&\SI{}{\kelvin}& Temperature in the vicinity of crystallization interface &$1683$\\
	\hline
	$T_{B,0}$&\SI{}{\kelvin}& Initial temperature of the meniscus &1688.6\\
	\hline
	$T_{bulk,0}$&\SI{}{\kelvin}& Initial temperature of the melt bulk&1704\\
	\hline
	$\rho_l$&\SI{}{\kilogram\per\meter\cubed}& liquid state density of Si &2570\\
	\hline
	$\rho_s$&\SI{}{\kilogram\per\meter\cubed}& solid state density of Si &2330\\
	\hline
	$C_p$&\SI{}{\joule\per\kilogram\per\kelvin}& Specific heat capacity of Si melt&1000\\
	\hline
	$\Delta\,H$&\SI{}{\joule\per\kilogram}& Latent heat of fusion &$1.79\times10^6$\\
	\hline
	$\phi_{s}$&\SI{}{\watt\per\meter\squared}& Heat flux into the solid crystal from the crystallization interface &$1.3\times10^5$\\
	\hline
	$\phi_{l}$&\SI{}{\watt\per\meter\squared}& Heat flux entering the meniscus &$4.6\times10^4$\\
	\hline
	$\epsilon_{m}$& -- &Melt emissivity \citep{gevelber1987dynamics}&$0.2$\\
	\hline
	$\sigma$&\SI{}{\watt\per\meter\squared\kelvin}$^{-4}$&Stefan-Boltzmann constant&$5.67\times10^{-8}$\\
	\hline
	$T_{env}$&\SI{}{\kelvin} &Temperature of the environment \citep{gevelber1987dynamics} &$1262$\\
	\hline
	$F_{mc}$&-- & Radiation view factor from free melt surface to the crystal surroundings &0.5\\
	\hline
	\multicolumn{4}{|c|}{\textbf{States and parameters exclusive to thermal model I}} \\
	\hline
	$T_{bulk}$&\SI{}{\kelvin}& Temperature of the melt bulk &$1704$\\
	\hline
	$T_{B}$&\SI{}{\kelvin}& Temperature at the base of the meniscus &$1688.6$\\
	\hline
	$Q_{mb}$&\SI{}{\kilo\watt}& Heat entering the meniscus   &7.3\\
	\hline
	$k_{cond,I}$&\SI{}{\watt\per\meter\per\kelvin}& conductivity of liquid Si at 1700K &57\\
	\hline
	\multicolumn{4}{|c|}{\textbf{States and parameters exclusive to thermal model II}} \\
	\hline
	$T_{bulk}$&\SI{}{\kelvin}& Temperature of the melt bulk including the temperature of the meniscus &$1704$\\
	\hline
\caption {Parameters/ states for thermal models I and II. The initial values are taken from \cite{RAHMANPOUR2017353}}.
\label{table:FIS}
\end{longtable}

\end{document}